# From the Earth to the Sun

Dieynaba Young*

*This year marks 360 years since the founding of the French Academy of Sciences. This is the story of the academy's daring 17th century expedition to measure the Earth-sun distance. A truncated version first appeared in National Geographic Magazine; the following is closer to the original draft, with updated information and references.*

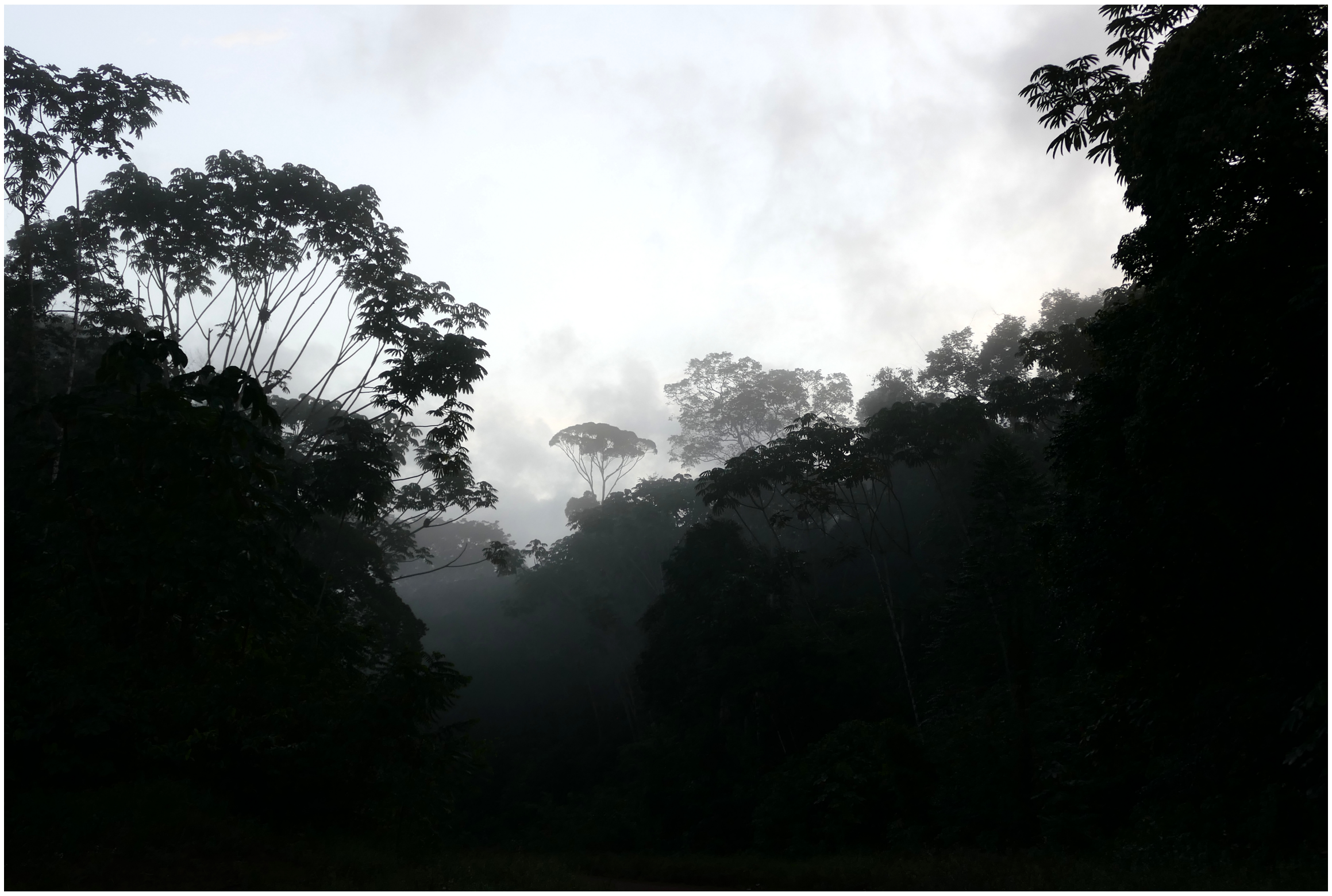

Fig. 1 - Morning Mist on the Amazon, French Guiana.
Bernard Dupont via Wikimedia Commons

On a drizzling day in May 1673, deep in the dense foliage of the Amazon rainforest, a scientist died. Known to historians only by his first name, Meurisse, he may have been cut down by disease or perhaps a fatal accident, but an accurate account of his death was never properly recorded. His sole companion was an astronomer named Jean Richer. Stricken ill beneath the unwavering heat and ceaseless downpour, Richer was fighting for his own life and unable to come to his colleague's aid.

The pair had been dispatched the year before from Paris, 4,400 miles away, by the French Academy of Sciences at the behest of their superior, famed astronomer Giovanni Cassini, to measure the distance between the Earth and the sun. Long after their deaths, the voyage to Cayenne, French Guiana would linger in the popular imagination as the world's first modern astronomical expedition.

* *Birkbeck, University of London; dyoungsend@gmail.com*

In many ways, the late 17th century journey was unremarkable. In fact, Richer and Meurisse would complete a dry run - an expedition to measure the heights of the tides on the coast of North America - two years before they set foot in the jungle; and there were several French expeditions shortly to follow - voyages to destinations as far flung as Gorée, Senegal and Quito, Ecuador.

But it would be this voyage that would collect data that, united with Cassini's trigonometric prowess, produced the first precise measurement of the astronomical unit - the distance between the Earth and the sun.

On January 11, 1667, roughly five years before the mission, astronomer Adrien Auzout stood in the meeting room of the sumptuous Bibliothèque du Roi in Paris, confidently reading aloud from his notes. Before a small assembly of men in long, thick wigs; he laid out a comprehensive program of astronomical research. " Right at the time the Academy is being conceived, they're *already* thinking about astronomical expeditions" explains McGill University Professor Nicholas Dew, leaning so far forward into his camera during our zoom call to emphasize his point that it's difficult to see the bright, book-strewn office behind him. "Auzout was the planner of this - [He had] the vision of using colonial trade networks to send observers to points around the globe, to conduct observations in astronomy, which, when collated and compared, could produce important data. "

Auzout's plan was wide-ranging and visionary. He recognized that certain astronomical questions, such as the distances to the planets and the sun, would require taking observations simultaneously in two distant locations - such as in Paris and a far-flung locale. Auzout argued for a voyage all the way to Madagascar, where the East India Company was expected to establish operations, and the proximity to the equator would allow astronomers to take key observations.

Measuring from a spot close to the equator would greatly reduce the effects of refraction. Atmospheric refraction describes the bending of a ray of light from a star (or any other heavenly body) as it passes through the Earth's atmosphere. The effect distorts the heights of stars, in particular those that are close to the horizon, making it difficult to take precise measurements.

As the men listened, the sounds and scents of a squalid, congested city must have wafted through the open windows. In late 17th century Paris, *fiacres* - horse-drawn carriages with large numbers painted in bright yellow on their sides - could be ordered by anyone for a reasonable fare. They sped along narrow, unpaved streets, heedlessly endangering wandering livestock and absent-minded pedestrians alike. Paris was a city known for raucous church processions, drunken merrymaking and open gun violence. At the strike of seven each morning, city officials marched down the wide boulevards, ringing bronze bells to wake residents, directing them to clean the filth that had accumulated in front of their homes or risk a fine.

In this hotbed of both commerce and intellectual fervor; a large, affluent population mixed freely with members of a forward-thinking scientific community.

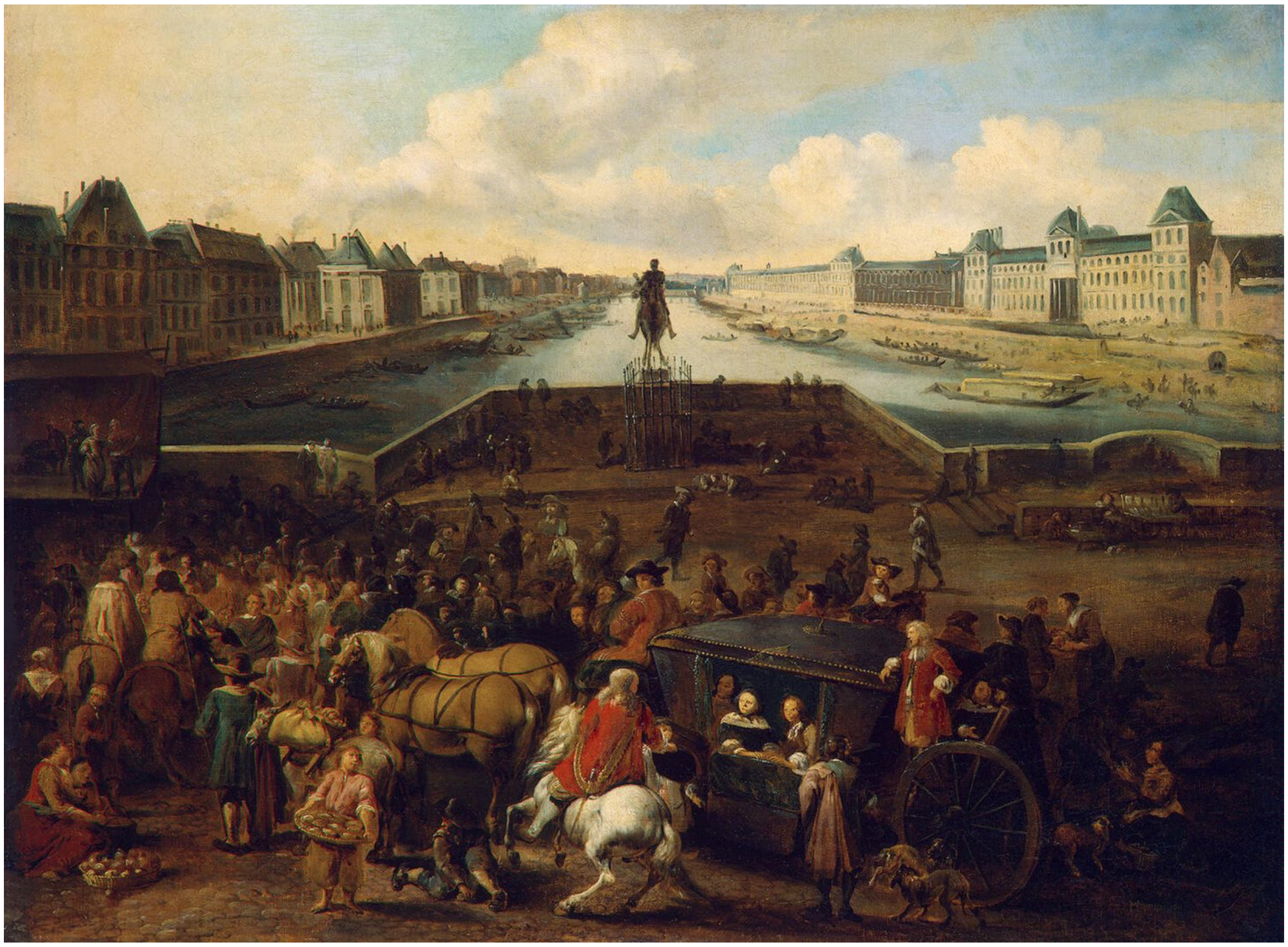

Fig. 2 - View of the Louvre from the Pont Neuf, Paris, with the Equestrian Statue of Henri IV.
The statue of Henri IV looms over a crowd gathered on the Pont Neuf, revealing the chaotic, bustling daily life of Parisians in the 17th century - a coach carrying noble passengers collides with a crowd of spectators absorbed in a theatre performance. Pedestrians rushing across the bridge intermingle with beggars and barefoot vendors hawking their wares. Unlike most bridges at the time, the Pont Neuf was stone, open-air and free of houses; making it a unique social hub in the city's center.
Hendrick Mommers, between 1666 and 1694, Louvre Museum

The most advanced astronomical and mathematical instrument makers in the world were steps away, their workshops concentrated in the Faubourg Saint-Germain district and on lle de la Cité. Far across town, at the edge of the countryside, near the Port-Royal Abbey, construction was beginning on the site that Louis XIV had approved for a new astronomical observatory, the first in France.

In the midst of all that excitement, founded in December of 1666 in the heart of the city, the new Academy of Sciences had begun to meet.

Founded by Jean-Baptist Colbert, the King's minister of finance, the academy had been created to carry out a range of scientific investigations - in the fields of geodesy, hydrography, navigation, astronomy, mathematics, optics, botany and more. All expenses would be paid by the royal treasury; but expeditions would require Colbert's express approval, and members would meet in the Bibliothèque du Roi, on rue Vivienne, just a block from his palatial townhouse.

Colbert had laid out the program carefully - the men would meet twice a week for anywhere from two to five hours  - mathematical sciences on Wednesdays, natural philosophy on Saturdays. Members were required to attend both days, but only months after meetings begun, recalcitrant scientists had begun breaking that rule, attending only on the days that focused on their areas of interest and expertise.

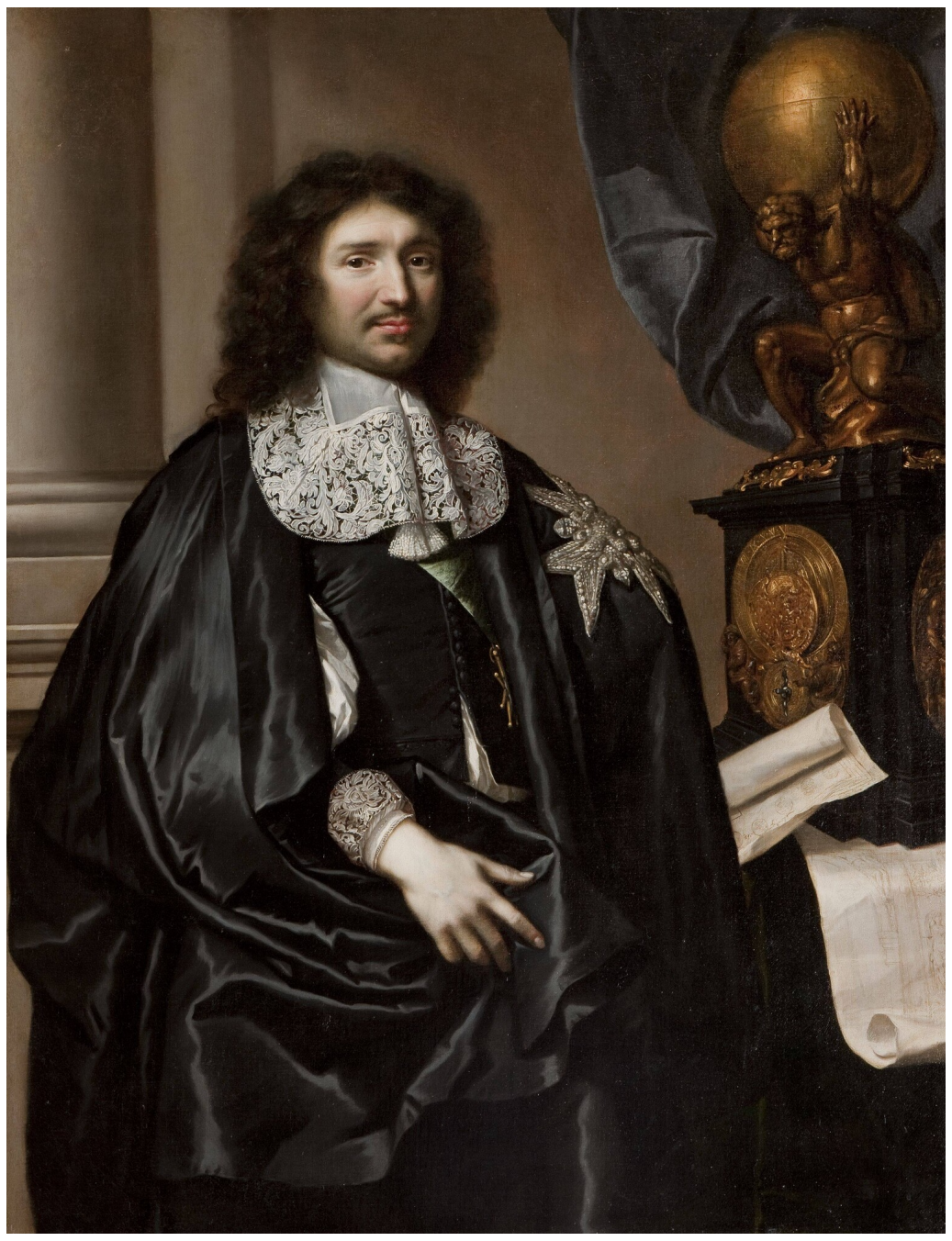

Fig. 3 - Jean-Baptiste Colbert.
Claude Lefèbvre,1666, Palace of Versailles Collection

Working directly with the academicians were the élèves, or apprentices, then, on a level below the apprentices, various assistants and outside contractors. These marginal positions were paid stipends, and sometimes, at least in the case of Meurisse, living expenses as well. For instance, later, on January 30 1670, Richer was paid 2500 livres and Meurisse were paid 1200 livres respectively as advance compensation for the coming year - they were already being described in Academy minutes as *mathematicians designated to go to Cayenne to make astronomical observations of utility to navigation.* Considering all that was to come, the pair probably deserved a bit more.

Two years after Auzout's speech, in April 1669, Giovanni Domenico Cassini arrived in Paris after accepting a personal invitation from the king. He would swiftly become one of the Academy's most illustrious figures.

"Cassini was 44 years old when he set off for Paris - a bachelor with a carriage full of astronomical instruments," explains Gabriella Bernardi, author of *Giovanni Domenico Cassini: A Modern Astronomer in the 17th Century*.

As the Academy continued to prepare for an astronomical expedition to the equator, the scientists shifted their focus from Madagascar to Cayenne. This French settlement was a shorter distance away, and the Academy had to act quickly to catch a noteworthy event: In the fall of 1672, Mars and Earth would be at their closest points to each other in 15 years.

Cassini realized that precise observations of Mars during this time from two separate locations could be used to calculate the parallax of the planet - a measurement of Mars' apparent difference in position as seen from the two observing sites. This key measurement could then be used to work out the distance from the Earth to the sun, making the close approach of Mars an opportunity that mustn't be missed.

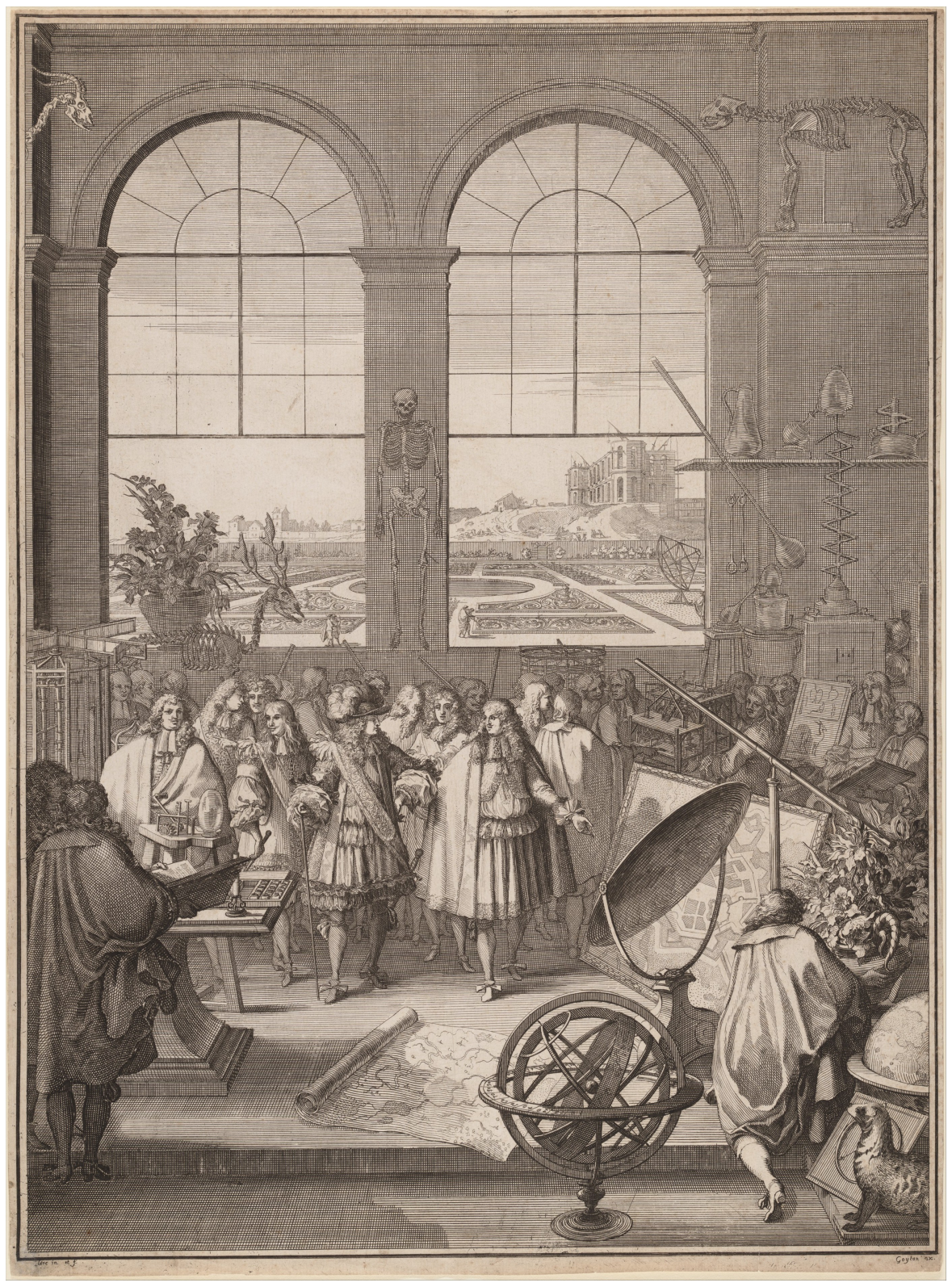

Fig. 4 - Louis XIV Visiting the Royal Academy of Sciences. Louis XIV is accompanied by the academy's leading figures, including Jean-Baptiste Colbert and Giovanni Cassini.
Sébastien Leclerc, 1671, Metropolitan Museum of Art

Cassini's idea was to calculate the parallax of Mars and use that figure, in conjunction with scientist Johannes Kepler's newly formulated third law to calculate the distance from the Earth to the sun.

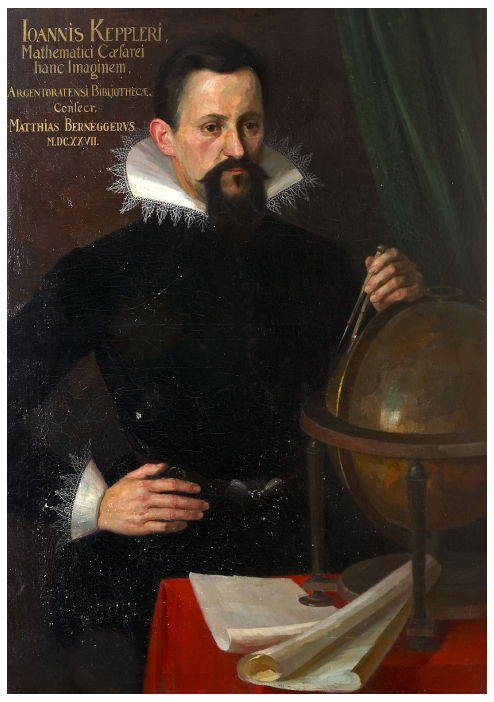

Fig. 5 - Johannes Kepler.
August Köhler,1910, Kepler Museum

"The mathematics was not particularly new -the general formula for the cubic equation has been known for a century. What made the difference was the use of Kepler's law, formulated in 1619, [and] the idea of measuring Mars at opposition . . . previous attempts since antiquity were based on much less reliable methods and obtained extremely inaccurate values." Bernardi explains.

As long as humankind has gazed up in wonder at the sky; there have been attempts to measure the distance to our closest star. Scientists in antiquity, particularly Eratosthenes and Ptolmey, had produced estimates that varied significantly, often greatly underestimating the true distance.

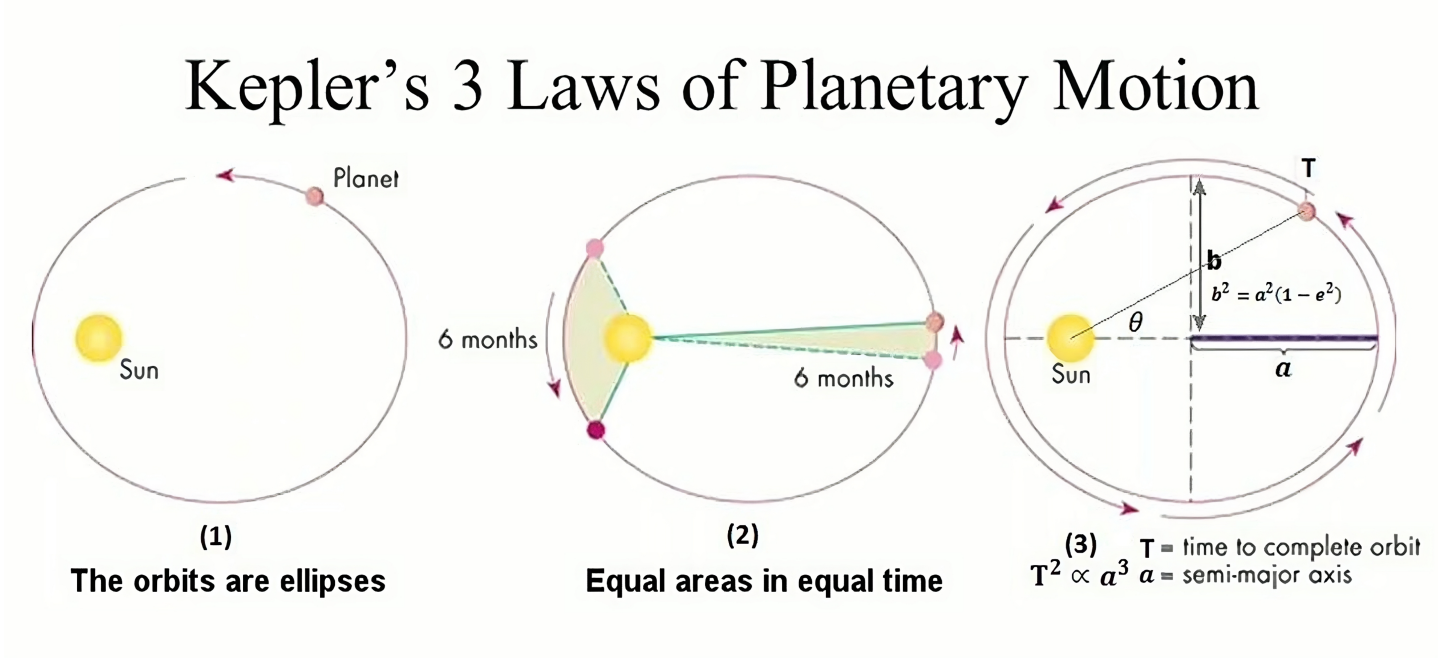

Fig. 6 - Illustration of Kepler's laws of planetary motion: 1st- the law of elliptical orbits, 2nd- the law of equal areas and 3rd- the law of harmonies; which states that the square of a planet's orbital period is proportional to its mean distance from the sun, cubed.
Spacenasa via Wikimedia Commons

By the 1670s, aided by newly developed technology, Cassini was determined to find the answer once and for all. Recent technical breakthroughs would, to a great degree, differentiate this attempt at measuring the Earth-Sun distance from all those that had come before. Historian John Olmsted explained in *The Scientific Expedition of Jean Richer to Cayenne (1672-1673)*:

> Between 1650 and 1670 three great inventions, the pendulum clock, the filar micrometer, and the application of telescopes to the graduated circles of astronomical instruments in place of the conventional sights or pinnules, combined to effect a revolution in observational astronomy. An entirely new standard of accuracy and refinement became possible. The stimulus to observational astronomy was immediate. And it was particularly consequential in France. The program of astronomical work which the Academy of Sciences early laid out for itself was based on a clear recognition that an entirely new degree of accuracy had become possible.

Inhabiting the second floor of the Paris Observatory, surrounded by cutting-edge instruments, Cassini worked on the problem unrelentingly. "He had no hobbies," says Bernardi. "From his diary emerges a man completely devoted to his profession."

Richer and Meurisse spent time working alongside Cassini, to prepare for the joint observations they'd have to make when they were thousands of miles away. As lowly apprentices, the pair knew that they were embarking on a perilous journey. "Anyone who's been sent on these ships in this period, they're all on a lower status level" Dew explains. "The dangerous, scary, long-distance travel is done by the lower grade, lower paid people”

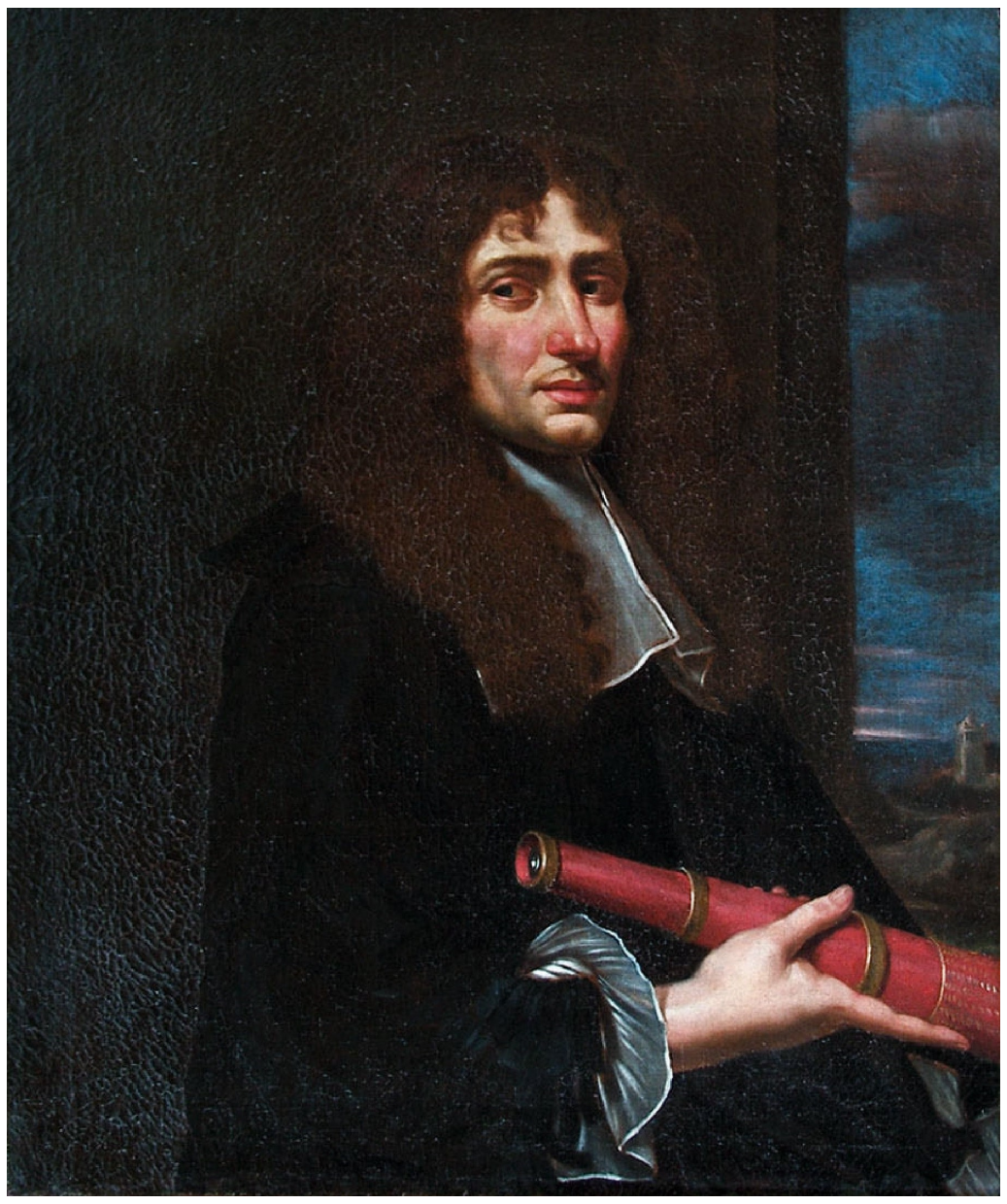

Fig. 7 - Giovanni Domenico Cassini.
Unknown artist, before 1673, Biblioteca Aprosiana

Traveling first to the French port of New Rochelle, Richer and Meurisse disembarked and spent three months in intense preparation - methodically testing and calibrating their instruments - an octant, a quadrant, several telescopes of various sizes, and a few pendulum clocks.

They set sail for Cayenne on February 8, 1672, on a merchant vessel - possibly an empty slave ship, on its way to Senegal. Gazing up from the ship's deck one evening during the passage over; Richer made detailed observations of a comet; two long, glowing tails streaking across an inky-black sky.

Cassini had given Richer several objectives, and he listed them carefully at the beginning of his 1679 mission report: He was to measure the positions of the southern stars, the heights of the tides, and the duration of twilight. He was to make observations of the solstices and equinoxes, and take detailed notes on the movements of Venus, Mars, and Mercury. He and Meurisse were expected to take barometric measurements and keep an eye out for unusual flora and fauna. Privately, and unofficially, Cassini slipped him special instructions to observe the eclipses of Jupiter’s moons and the occultations of a handful of fixed stars.

There was one small, final thing, not a main assignment of the expedition, in fact almost an afterthought. Richer was to measure the length of the simple pendulum - a clock that had an oscillation period of exactly one second.

The pair arrived in Cayenne on April 22nd, 1672.

*Fert aurum industris* - Work brings wealth. Whoever coined Cayenne’s official motto must have had a grim sense of humor. The tiny, desolate settlement could not have been an encouraging sight. Visited by only two or three ships a year, the ‘island’ was separated from the rest of Guiana by the narrow 11-mile Mahury estuary on one side and the slender Cayenne river on the other.

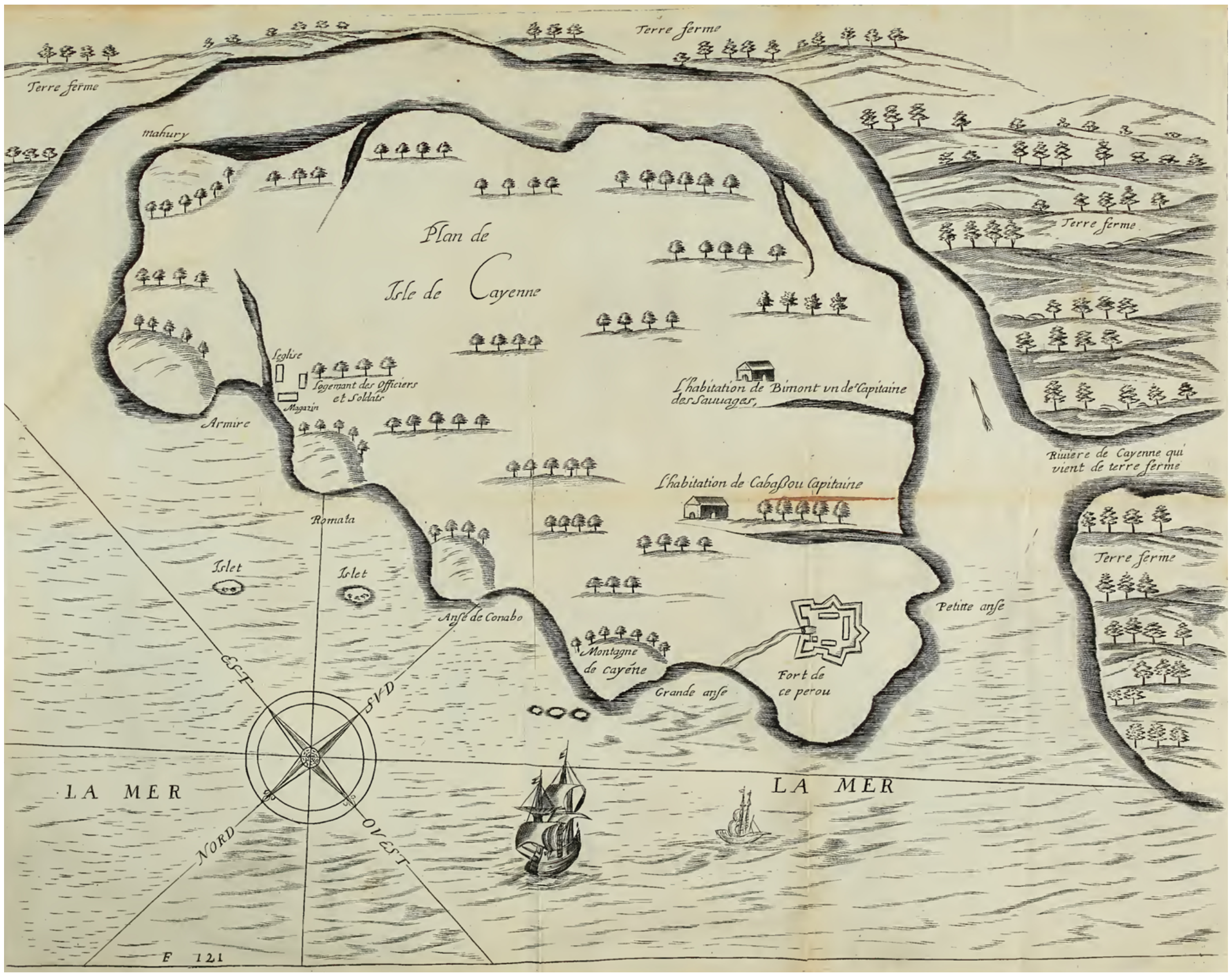

Fig. 8 - Anonymous 1652 map (published in 1654). Twenty years before Richer's arrival, Fort Cépérou's wooden walls were replaced with stone.
*Relation du voyage des François en l'Isle de Cayenne*, Jean de Laon.

As they stepped off the boat, Richer must have realized that they had chosen the most unpleasant time of year to arrive. In the Amazon, late April is near the height of monsoon season, oppressively humid and thick with mosquitoes. The sheets of rain fell on them mercilessly, flooding the river and yet providing no relief from the sweltering heat.

At the base of the settlement loomed Fort Cépérou, a bleak, lonely structure, rebuilt into stone from wood after the most recent attack by the native population; signifying the French colonists' stubborn determination to stay. A short walk from the fort was the King's Store, a kind of general store that served the whole settlement yet oftentimes had nothing on the shelves.

There was also a modest Jesuit church and mission house. A 1685 account, cited in Catherine Losier's *Supplying Cayenne Under the Old Regime: Archeology and History of Commercial Networks*, describes it as a dwelling occupied by four fathers and a brother, along with 82 enslaved African persons - 32 men, 23 women, and 27 children - to work the Jesuits' crops and tend to their livestock. Enslaved Africans made up roughly 85 percent of the settlement.

And then, there were the Kalina. The Indigenous tribe, formally referred to as the Galibí, had resided in the Cayenne region for over two thousand years before Europeans arrived. As one settler, Paul Boyer, would write after a visit around 1654: "All the Galibí could think about was how to be rid of the French."

Past interactions between the two groups had been troubled. Less than 30 years before Richer's arrival, in 1644, a haughty Parisian nobleman named Charles Poncet de Brétigny had arrived in Cayenne with roughly three hundred men and declared himself 'king' of the colony. He had used an iron cattle brand bearing his name on the Kalina who displeased him, tried to force them to wear clothes, and abducted Indigenous women, confining them to his quarters - including the beautiful wife of the chief.

Within a year, a tribesman had sunk an ax into Bretigny's skull, the opening of a blood-soaked ambush that left only a handful of Frenchmen alive in a settlement that had been burned to the ground.

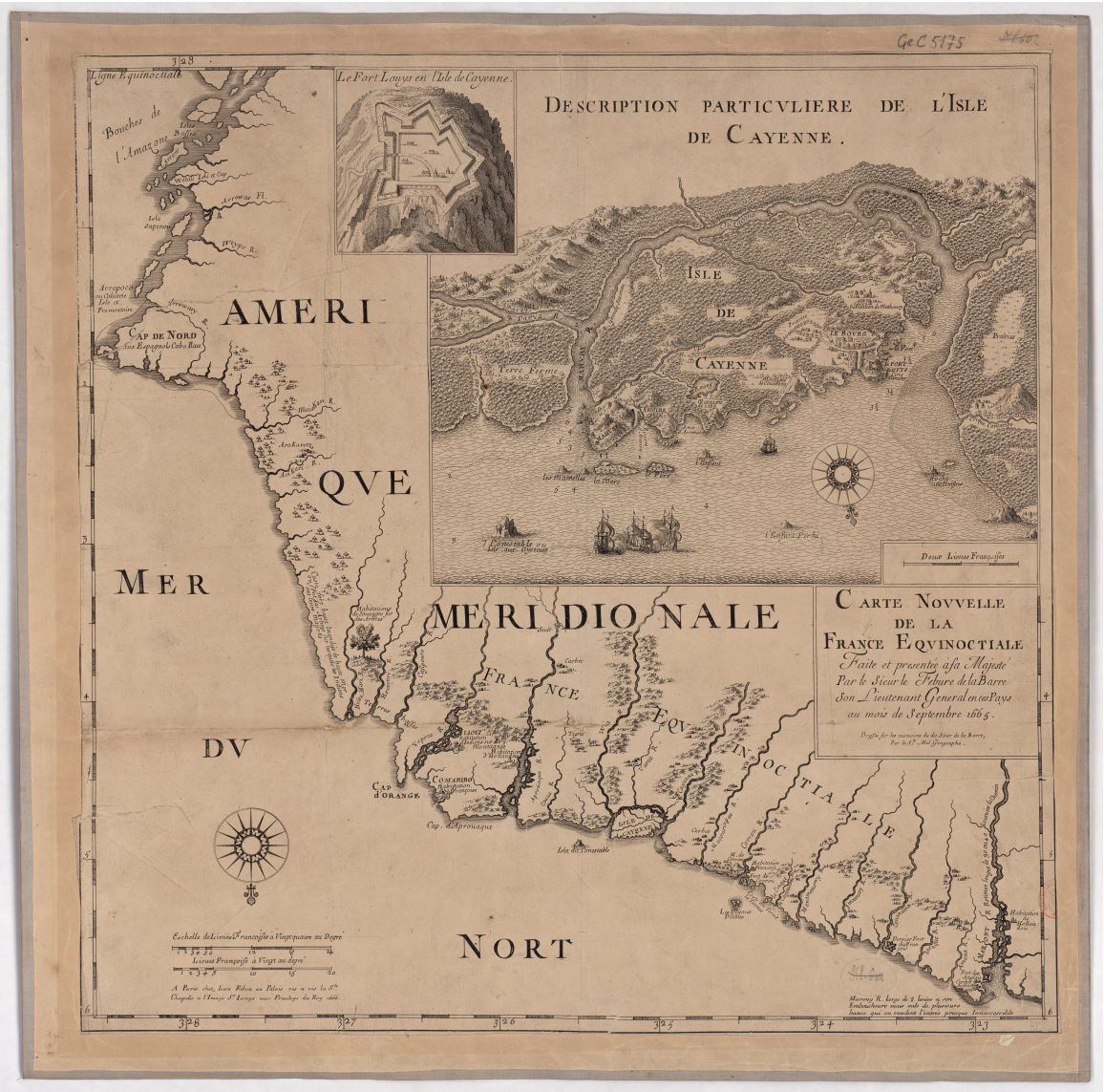

Fig. 9 - Carte Nouvelle de la France Equinoctiale. Cayanne was a tiny settlement on the northeastern coast of South America, known as the 'Wild Coast'.
Joseph-Antoine Lefebvre de la Barre,1666, Bibliothèque nationale de France

The French didn't only have the Kalina to worry about. The Dutch managed to capture the colony a decade after Bretigny's reign, only to be forced out by fresh French troops in a surprise attack. French settlers were then driven out by the British in 1667, wrenching back control of the colony - only four years before Richer's feet touched the shore.

For Louis XIV, Guiana provided a strategic position on the continent from which his country - squeezed out by the Portuguese further south, and eventually hemmed in by the Dutch to the east (Surinam) and the British to the west (British Guyana) - could gain a foothold on the continent.

There was something else that was tantalizing about Cayenne, often divulged in a whisper - *El Dorado*. Every European fighting to control the region believed that the fabled city of gold was hidden somewhere inland from the northern coast of South America; and whoever controlled Cayenne would have a direct route to the riches.

*A path that leads to peril and mishap* Walter Raleigh had written in his 16th century poem *A Farewell to False Love*. He might have been describing the European exploration of the amazon. El Dorado had been the lure that had drawn many, including Raleigh himself, across the Atlantic on a quest for treasure of unimaginable value.

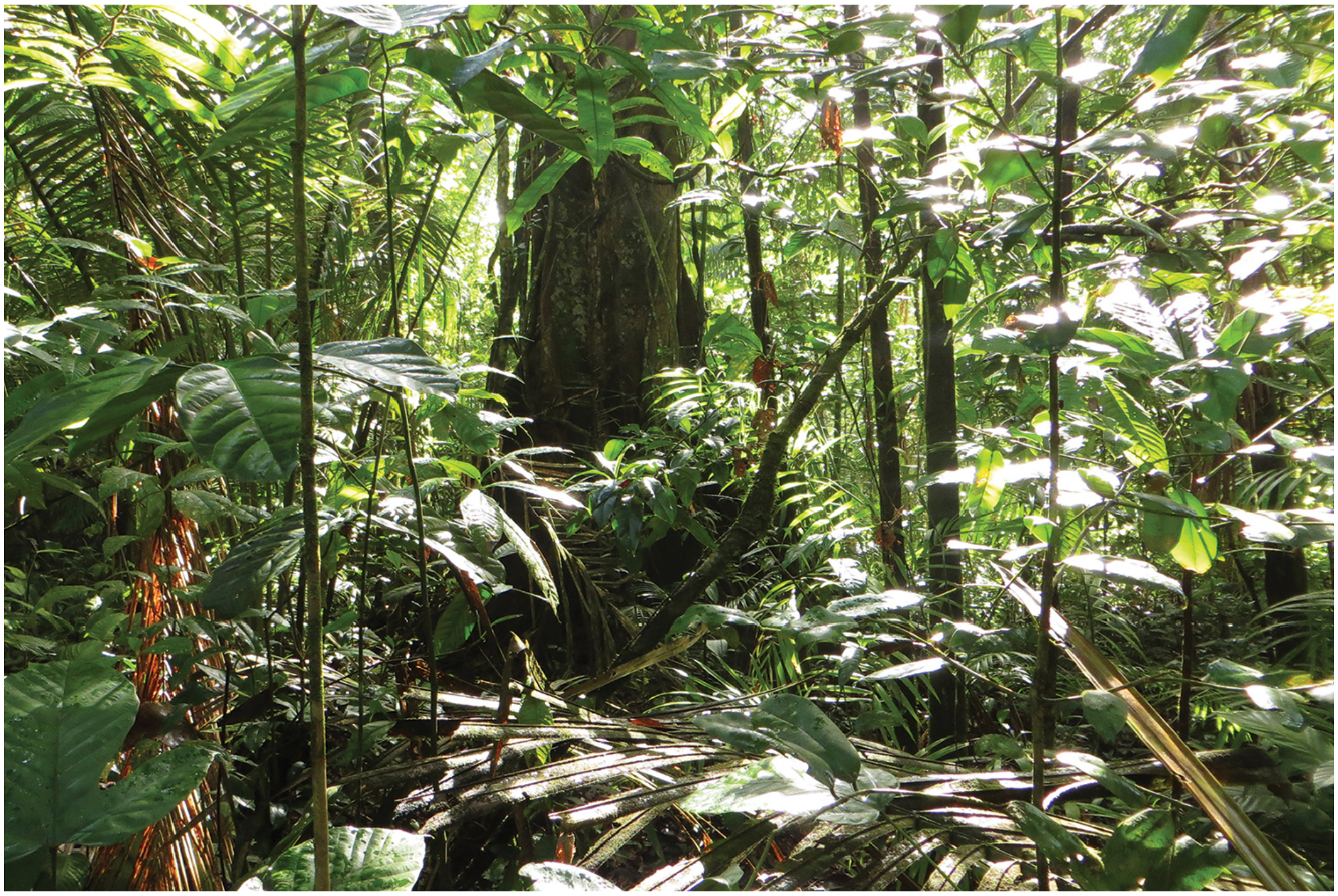

Fig. 10 - Lowland tropical rainforest, French Guiana. Many believed El Dorado was hidden somewhere deep in the uncharted interior.
Ythier E via Wikipedia Commons

Raleigh too had searched for the fabled kingdom, ruled by a tribe that worshipped the sun, on the banks of a lake overflowing with viridescent gemstones. In the end, countless Europeans perished in the jungle in pursuit of wealth.

Now Jean Richer had arrived at the same impenetrable wilderness, in desperate search of something worth more than gold.

Just steps away from the settlement, across the thin, narrow river that gave Cayenne island it's name, lay the rest of Guiana - dense primeval rainforest, containing species of plants and animals not found anywhere else in the world - an environment so completely alien, so different from the cobbled streets of Paris; it's hard to imagine which species of wildlife would have caught the explorers' eyes first - the anteaters, iguanas or spider monkeys? Were they astonished by glimpses of speckled jaguars or bright green parrots?

Academy records indicate that Richer and Meurisse took pages of detailed notes on flora and fauna, almost all of which have been lost to time.

Plunging into the interior would have yielded glimpses of bright blue poison dart frogs, titan beetles - the largest of the species; and *Anelosimus eximius*, social spiders that constructed their communal webs up to 20 feet across. *Morpho rhetenor*, Guiana's famous, enormous iridescent blue butterflies are a common sight in the area, along with six-foot long boa constrictors and emerald anacondas, reaching up to thirty feet.

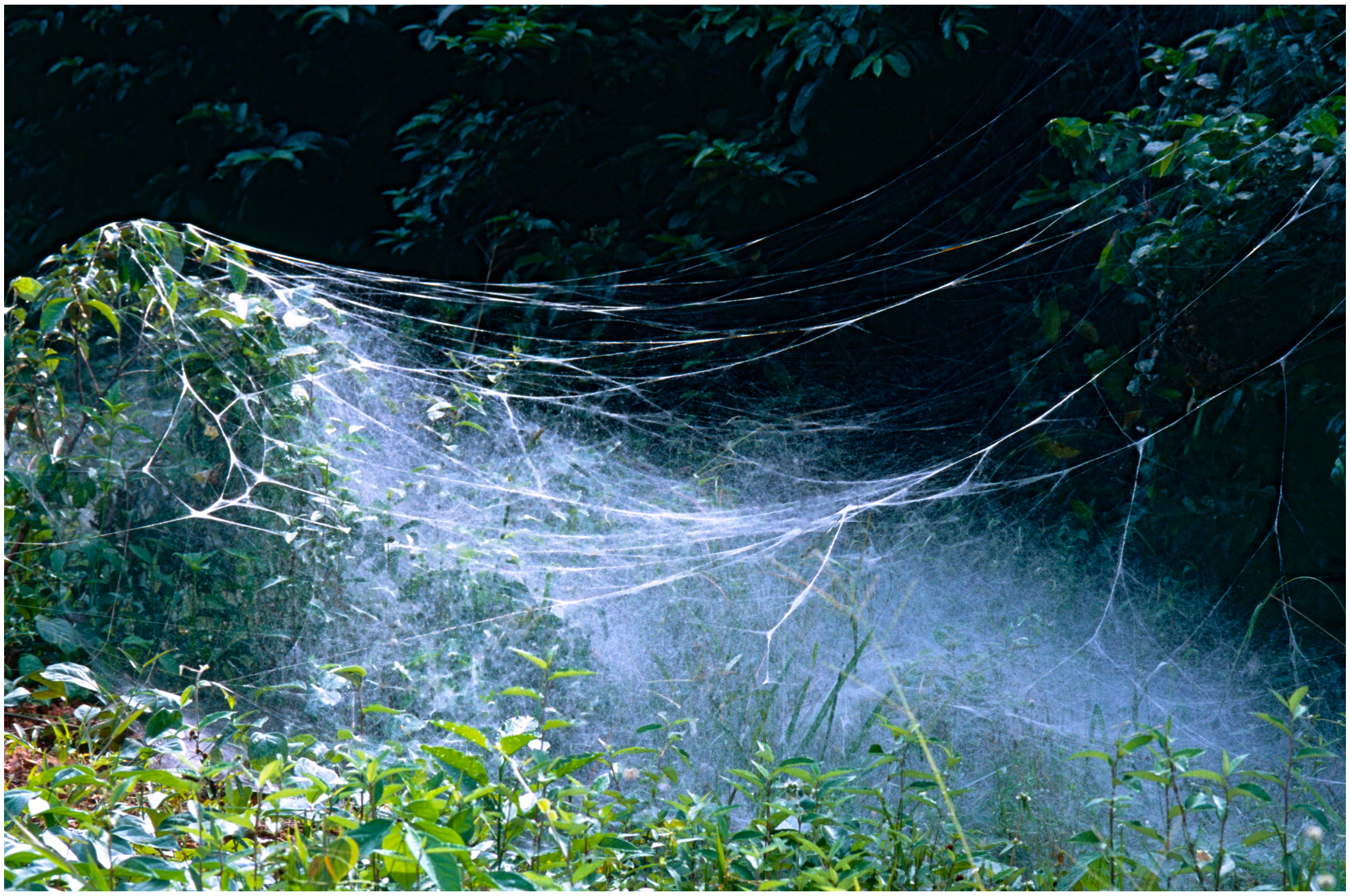

Fig. 11 - Anelosimus eximius web, French Guiana.
Bernard Dupont via Wikimedia Commons

Immediately upon arrival, Richer began scouting the jungle for the best place to build their observatory. Locating a spot after a couple of weeks, the two men recruited Indigenous workers and built a structure composed of branches, tree bark, and palm tree leaves, with a sizeable hole in the roof for their telescopes.

Sometime before mid-May the observatory was finished. The pair then set about calibrating and testing their instruments for a third and final time. Richer's first observation was on May 14. He calculated the height of the North Star. It was a promising start to what would be a very challenging mission.

The rain was merciless, and Richer wrote to Cassini of not being able to take observations for several days at a time because of the inclement weather. "Almost not a day has passed without rain since our arrival"

At some point, so many ants crawled into their pendulum clocks; that the creatures jammed the delicate machinery of cogs and wheels, causing at least one clock to stop completely.

When not making astronomical observations; Richer explored the jungle.  He captured, killed and dissected a peccary and came face to face with an electric eel. He noted in the appendix of his 1679 report:

> I was even more surprised to see a three to four foot fish, similar to an eel, as fat as a leg, which the fishermen call a conger, that by simple touch with a finger or the tip of a stick, so numbs the arm and that part of the body closest to it that one remains about 15 minutes without being able to move. ... I was witness to this effect and I felt it, having touched the fish with my finger one day . . .

In his second letter to Paris, dated July 20, he complained of a shortage of food and asked whether Cassini had ever received his first letter, sent May 4. He had no way of knowing that Cassini had received it, and would receive the July letter as well.

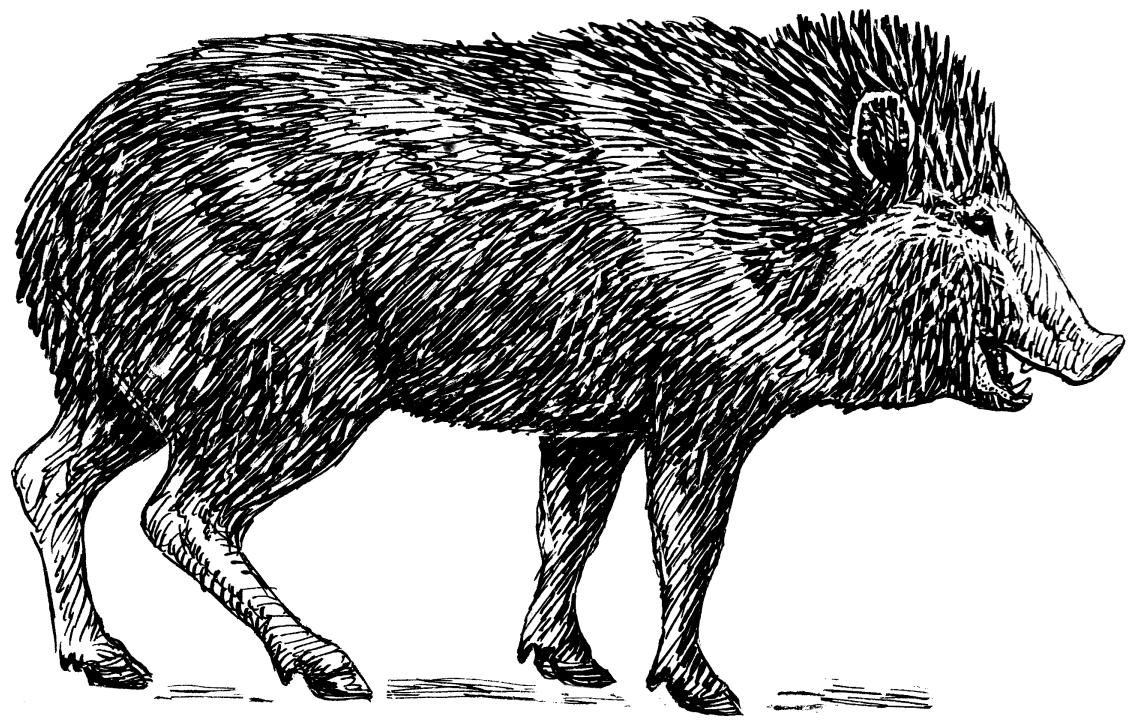

Fig. 12 - Peccary. During his time in the jungle, Richer dissected a peccary.
Pearson Scott Foresman via Wikimedia Commons

At first glance, Richer's struggles with provisions are hard to comprehend. Deep in the jungle, game is plentiful. The waters surrounding them would have been teeming with diverse species of catfish, oysters, red snapper, glass knifefish, piranha (roasted, they're a delicacy) and giant prawns. Just steps onto the shore, the same animals that the local green anacondas feasted on - capybaras, tapirs, caimans - should have provided abundantly. There were edible plant life - bananas, avocado, cassava plants; pineapple, lychee and mango trees.

Yet the pair relied almost solely on supplies from home - cloth wrapped packets of cured meat, flour, Bordeaux wine, coffee, cheese - supplies that were seldom replenished by passing ships. "The Europeans want to eat what they're used to eating . . .they're thinking - 'we have to have bread, we have to have wine' ". Dew says.

Along with foodstuff, they probably carried scanty supplies - candles, a few earthenware bowls and perhaps a run-of-the-mill red copper saucepan. Tossed into their sacks might have been the familiar dark-green tinted glass bottles manufactured at the time in a factory in the Gironde region of southwestern France; and thick rolls of calfskin parchment. The slowness of mail and the rareness of passing ships meant that the apprentice's questions and concerns would go unanswered. Effectively, whatever happened now, Richer and Meurisse were on their own.

Finally in October 1672, the rainy season stopped - just in time to observe Mars. Richer measured the planet and nearby stars over the course of multiple weeks.

4,400 miles away, Cassini, accompanied by Danish astronomer Ole Rømer, also made his measurements at the agreed upon times, peering out the window of the Paris Observatory.

Meanwhile, in London, astronomer John Flamsteed of the Royal Society was also measuring the parallax of Mars to determine the distance to the sun, cleverly observing Mars once early in the evening, waiting several hours for Earth to rotate, then measuring again. His final calculation of the distance would be close, but not quite as precise as Cassini's.

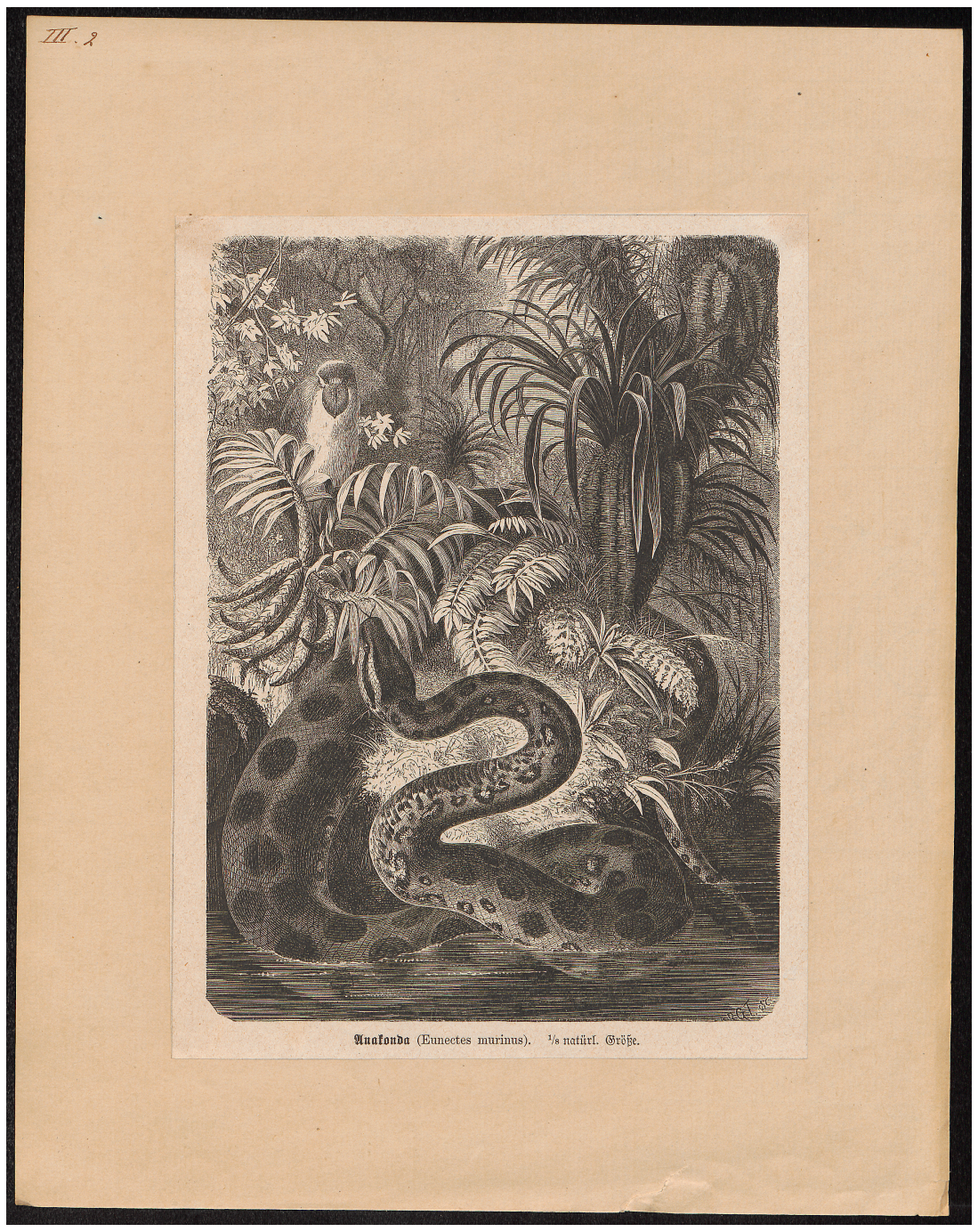

Fig. 13 - Eunectes murinus (green anaconda).
Iconographia Zoologica, 1700- 1880, Special Collections of the University of Amsterdam

In the spring of 1673, Meurisse died under mysterious circumstances - possible causes include yellow fever, malaria, pneumonia or even severe malnutrition. After all his valuable work for the academy, his death would be noted once in the record; then never mentioned again. "When people left European shores to go on these far [voyages] - long distance, across the ocean - they assumed that you might die" Dew says.

Richer, now alone, felt too ill to stay on. He sought out specimens to carry back to Paris - capturing a live crocodile and chaining it up in the hold of the ship; plucking an unspecified plant that he would carefully dry and present to academy members. Wracked with illness, he boarded the ship and departed Cayenne on May 25, 1673. On the long journey home, the crocodile would die of starvation, but Richer would recover.

In 1679 his official mission report, *Observations astronomiques et physiques faites en l'isle de Caienne*, was released. Aided by Richer's data, Cassini could finally make his calculations, releasing the results in a 1684 publication. He declared that our sun, which looked so close, was actually 87 million miles away.

Word of the expedition and revelation at the sheer size of the solar system spread rapidly, in large part due to the popular writings of Bernard le Bovier de Fontenelle, who had been made an official secretary of the Academy of Sciences in 1697. Fontenelle wrote about science in a unique, novelistic style; encouraging interest in expeditions to flourish. His accessible, easy to understand books about physics and the history of the Academy of Sciences were widely available. The calculation of the astronomical unit would inspire widespread amazement and excitement, and in the years following the Cayenne voyage, numerous other expeditions were approved. For the first time, astronomy was a subject for the dinner table.

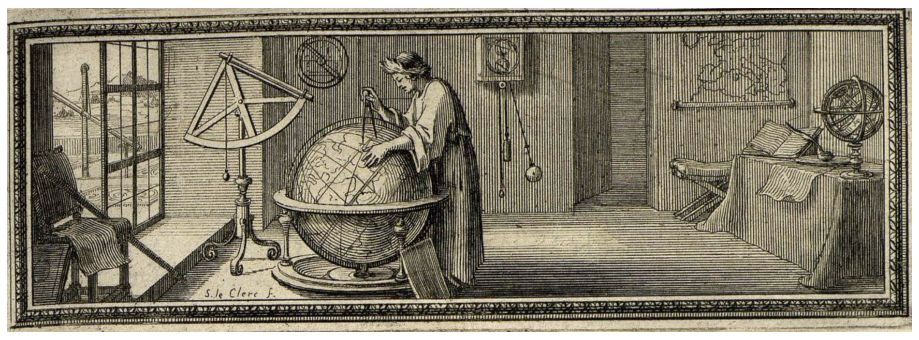

Fig. 14 - Jean Richer engaged in scientific work on the cover of his famous mission report.
Engraving by Sébastien Leclerc, Text by Jean Richer, *Observations astronomiques et physiques faites en l'isle de Caïenne*, 1679, Bibliothèque nationale de France

*Just think, Milord* . . . Voltaire scribbled in a letter to his friend Lord Hervey in 1740 *without the voyage and experiments of those sent by Louis XIV to Cayenne in 1672 … never would Newton have made his discoveries concerning attraction.*

Richer's simple pendulum observations, made almost as an afterthought, would also have long lasting implications.

While in the Amazon, the astronomer had also measured the length of a second with a simple pendulum and compared the results to his precisely calibrated clocks. He realized that something was off. The swinging pendulum seemed to produce a shorter second in Cayenne than it would in Paris.

Richer didn't fully comprehend it at the time, but this difference was due to the fact that there is slightly less gravity near the equator, where Earth bulges as it spins, causing the pendulum to measure a shorter second. Isaac Newton would puzzle out the reason some 15 years later, using Richer's measurements as evidence for his new theories of gravity.

Bernardi believes that the success of the voyage was due to Cassini's modern approach. "At the time, it was a complete innovation," she says. "Cassini was the first to understand that a regular plan of observation, in collaboration with many other colleagues, made it possible to tackle more difficult problems and achieve important results, just like big science does today."

Just as the trading vessel that carried him home quietly slipped away from the green shores of French Guiana, Richer would slowly slip into relative obscurity, the accurate calculation of the astronomical unit becoming almost wholly Cassini's triumph. Once safely back in France, he would break from the academy and move to a prestigious position assisting the famed military engineer Sébastien Vauban; then disappear from the historical record entirely.

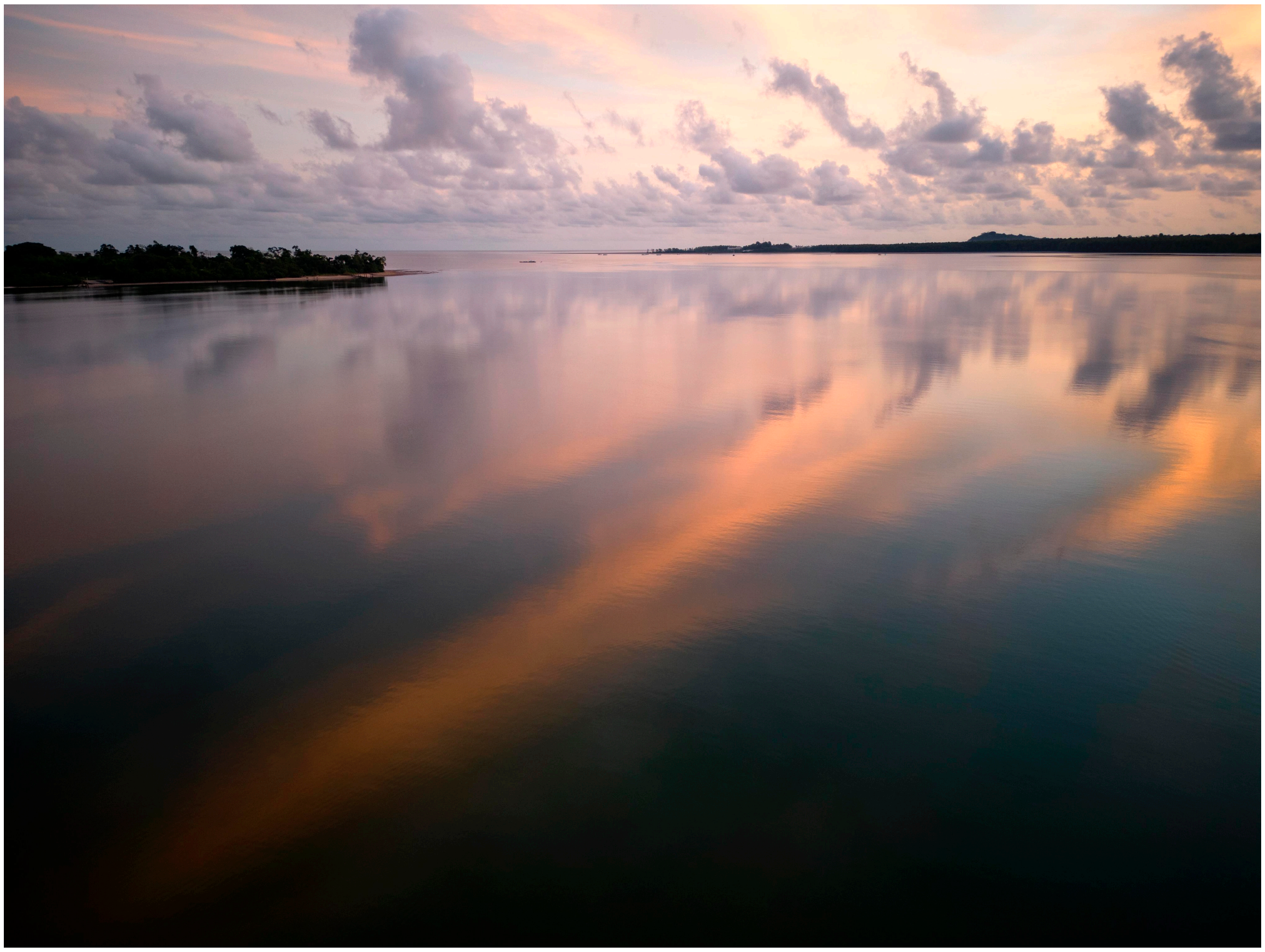

Fig. 15 - Dawn at the mouth of the Cayenne River.
Stephane Lesbats via Wikimedia Commons

*What is our life?* Walter Raleigh, fellow seeker, had asked at the beginning of his famous poem of the same name. *Our graves that hide us from the searching sun, are like drawn curtains when the play is done.* On the hot day Richer left the jungle, did he turn back and make a final, painful observation? The site of his comrade's death had also been the site of a tremendous intellectual achievement - not only for the French Academy, but for humankind. They had sought knowledge, not gold; their quest had ended in triumph; where so many before had fallen.

It would have been, once again, the height of monsoon season when Richer's ship sailed away. The river would have been close to overflowing; the severe flooding and the relentless, unceasing rain tormenting the handful of European settlers he was leaving behind.

Yet it’s possible, after all he'd suffered, that a hand-scrawled draft of his mission report was already tucked away alongside his last functioning, carefully calibrated pendulum clock; bob swaying back and forth, back and forth, marking time as faithfully as the rising and ebbing of the tide.